\documentclass[sigconf, nonacm]{acmart}
\usepackage{kotex}

\AtBeginDocument{%
  }

\setcopyright{acmlicensed}
\copyrightyear{2026}
\acmYear{2026}

\acmConference[SIGIR '26]{The 49th International ACM SIGIR Conference on Research and Development in Information Retrieval}{July 20--24, 2026}{Melbourne | Naarm, Australia}

\usepackage{booktabs}
\usepackage{tabularx}
\usepackage{array}
\usepackage[table]{xcolor}
\usepackage{soul} 
\sethlcolor{yellow!35} 
\usepackage{makecell}
\usepackage{siunitx}
\begin{document}
\author{Hyewon Choi}
\email{hyewon.choi@lgresearch.ai}
\affiliation{%
  \institution{LG AI Research}
  \city{Seoul}
  \country{Republic of Korea}
}
\author{Jooyoung Choi}
\email{jooyoung.choi@lgresearch.ai}
\affiliation{%
  \institution{LG AI Research}
  \city{Seoul}
  \country{Republic of Korea}
}
\author{Hansol Jang}
\email{hansol.jang@lgresearch.ai}
\affiliation{%
  \institution{LG AI Research}
  \city{Seoul}
  \country{Republic of Korea}
}
\author{Hyun Kim}
\email{hyun101.kim@lgresearch.ai}
\affiliation{%
  \institution{LG AI Research}
  \city{Seoul}
  \country{Republic of Korea}
}
\author{Chulmin Yun}
\email{chulmin.yun@lgresearch.ai}
\affiliation{%
  \institution{LG AI Research}
  \city{Seoul}
  \country{Republic of Korea}
}
\author{ChangWook Jun}
\email{cwjun@lgresearch.ai}
\affiliation{%
  \institution{LG AI Research}
  \city{Seoul}
  \country{Republic of Korea}
}
\author{Stanley Jungkyu Choi}
\email{stanleyjk.choi@lgresearch.ai}
\affiliation{%
  \institution{LG AI Research}
  \city{Seoul}
  \country{Republic of Korea}
}
\title{ARHN: Answer-Centric Relabeling of Hard Negatives with Open-Source LLMs for Dense Retrieval}


\renewcommand{\shortauthors}{Choi et al.}

\begin{abstract}

Neural retrievers are often trained on large-scale triplet data comprising a query, a positive passage, and a set of hard negatives. In practice, hard-negative mining can introduce false negatives and other ambiguous negatives, including passages that are relevant or contain partial answers to the query. Such label noise yields inconsistent supervision and can degrade retrieval effectiveness. 

We propose ARHN (Answer-centric Relabeling of Hard Negatives), a two-stage framework that leverages open-source LLMs to refine hard negative samples using answer-centric relevance signals. In the first stage, for each query–passage pair, ARHN prompts the LLM to generate a passage-grounded answer snippet or to indicate that the passage does not support an answer. In the second stage, ARHN applies an LLM-based listwise ranking over the candidate set to order passages by direct answerability to the query. Passages ranked above the original positive are relabeled to additional positives. Among passages ranked below the positive, ARHN exclude any that contain an answer snippet from the negative set to avoid ambiguous supervision.

We evaluated ARHN on the BEIR benchmark under three configurations: relabeling only, filtering only, and their combination. Across datasets, the combined strategy consistently improves over either step in isolation, indicating that jointly relabeling false negatives and filtering ambiguous negatives yields cleaner supervision for training neural retrieval models. By relying strictly on open-source models, ARHN establishes a cost-effective and scalable refinement pipeline suitable for large-scale training.
\end{abstract}




\maketitle


\section{Introduction}
\label{sec:intro}

\begin{figure}[t]
    \centering
    \includegraphics[width=\columnwidth]{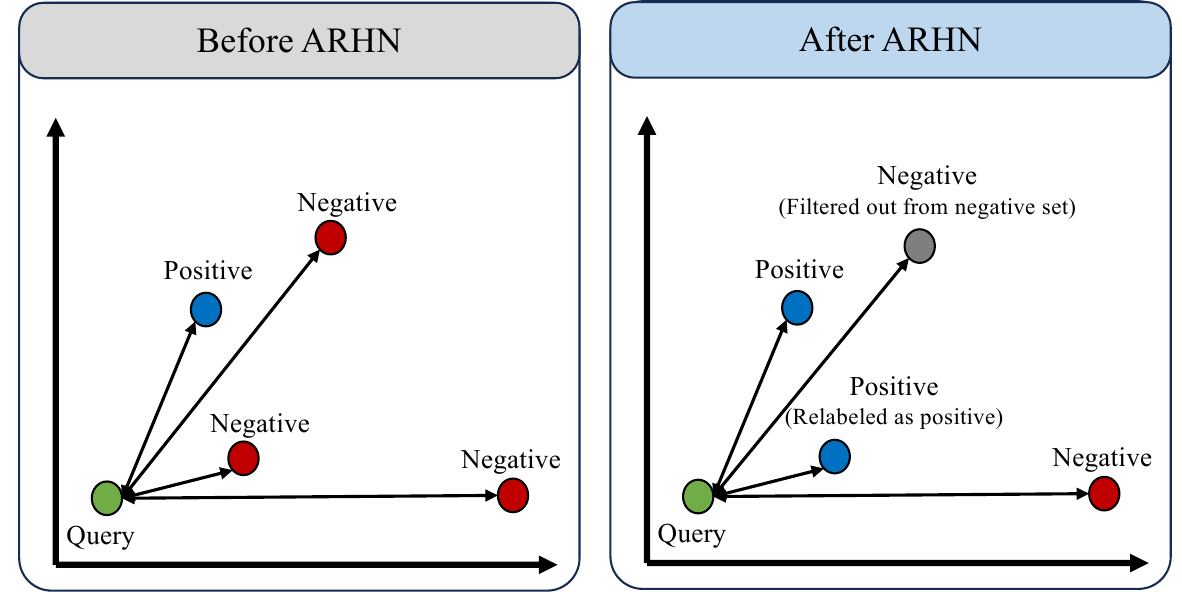}
    \caption{\textbf{Intuition of ARHN.}
(Left) Conventional hard-negative mining can include answer-bearing documents as negatives, introducing \emph{false negatives}.
(Right) Using extracted \emph{answer snippets}, ARHN relabels answer-bearing negatives as positives and filters ambiguous negatives to refine supervision.}

    \label{fig:arhn_before_after}
\end{figure}
In recent Information Retrieval (IR), dense retrieval has become a core component of various downstream tasks—including open-domain QA, RAG, and document retrieval—by mapping queries and documents into a shared embedding space and retrieving relevant documents based on similarity \cite{lewis2020retrieval,zhao2024dense,zeng2024unsupervised}.
Dense retrievers are typically trained with contrastive objectives that pull query–positive pairs together and push query–negative pairs apart, yielding a representation space where nearest-neighbor search recovers relevant documents. In RAG pipelines, this ranking determines which documents are provided to the generator as evidence. As a result, retrieval quality directly bounds answer quality: missing or mis-ranked evidence can immediately degrade the generated response \cite{zhao2024towards,wang2025astute,yan2024corrective,cuconasu2024power}. 
To produce reliable rankings that support downstream grounding, dense retrievers depend heavily on large-scale training data, and large query–passage pair datasets serve as a crucial foundation for stably shaping the representation space of contrastive-learning-based dense retrievers \cite{zhan2021optimizing,karpukhin2020dense,xiong2020approximate,qu2021rocketqa,moreira2024nv,rajapakse2024negative}.

To improve dense retrieval performance, substantial effort has been devoted to extracting and leveraging hard negatives.
In contrastive training, negatives are typically sampled from documents that are not annotated as relevant to the query. Hard negatives are a subset of these negatives that are lexically or semantically similar to the query. Hard negatives are widely used because they yield a more informative discriminative signal during training \cite{rajapakse2024negative,cai2022hard,kalantidis2020hard,yang2024trisampler}.
However, hard-negative pools used in large-scale datasets often include \emph{false negatives}, documents that are in fact relevant or contain answer evidence but are treated as negatives due to missing annotations. 
Such contamination corrupts the contrastive objective and can substantially degrade retrieval performance \cite{wang2024mitigating,xiong2020approximate,rajapakse2024negative,thakur2025hard}. 
This problem is exacerbated in open-domain IR, where exhaustive relevance labeling is infeasible and unlabeled documents are routinely used as negatives by default \cite{ni2025diras,cohen2024indi}.

Conventional hard-negative mining can introduce false negatives because it emphasizes difficulty without explicitly checking answer relevance. In practice, negatives are often mined from the top-ranked results returned by the retriever being trained, or from high-scoring candidates produced by BM25 or a re-ranker. However, these pipelines typically do not verify whether a mined candidate contains the answer or provides sufficient evidence to support it. Consequently, the negative pool may include passages that directly state the answer or allow the answer to be inferred from partial evidence. Training on such false negatives forces the model to repel semantically similar documents, including genuinely answer-bearing passages, which distort the embedding geometry and can destabilize contrastive learning. Therefore, constructing hard-negative data should consider not only negative difficulty but also the likelihood that a candidate is answer-bearing or relevant.

In this paper, we argue that hard-negative pools often contain false negatives because documents are mined and ranked without verifying whether they contain an answer to the query. To address this problem, we propose ARHN (Answer-Centric Relabeling of Hard Negatives), a data reconstruction framework that uses open-source LLMs to assess answer support among mined hard negatives and revise their labels.
Concretely, ARHN (1) extracts an answer snippet from the document text for each query–document pair, (2) re-ranks candidates in a listwise manner based on how directly the extracted snippet answers the query, and (3) either promotes (relabels) hard negatives to positives or removes (filters) them from the negative pool according to their ranking positions.

Figure~\ref{fig:arhn_before_after} illustrates the behavior of ARHN. Before applying ARHN, the standard hard negative mining process may include negative samples that answer the query more directly than the positive document, and incorporating them into training can induce supervision that pushes false negatives away as negatives. After applying ARHN, based on in-document evidence (answer snippets), ARHN (i) promotes hard negatives that provide more direct answers to the query to positives (relabeling) and (ii) removes ambiguous negatives that contain partial answer cues and thus are overly similar to positives from the negative set (filtering), thereby constructing a training signal centered on true negatives.

We evaluate dense retrievers fine-tuned on ARHN-refined data on the BEIR benchmark, and compare three settings: (i) relabeling only, (ii) filtering only, and (iii) their combination \cite{thakur2021beir}. Empirically, combining the two strategies consistently outperforms either single strategy and achieves competitive performance relative to prior methods. Moreover, by relying on open-source LLMs rather than commercial APIs, ARHN provides a cost-effective and scalable data refinement pipeline.

The contributions of this paper are as follows:
\vspace{-0.3em}

\begin{itemize}
\item We propose ARHN, a framework that leverages open-source LLMs to combine answer snippet extraction with listwise re-ranking, and present a data refinement strategy that integrates false-negative \emph{promotion (relabeling)} with \emph{filtering} of borderline samples.
\item Using open-source LLMs of varying scales, we systematically analyze the reconstruction process itself, and quantitatively characterize how changes in the \emph{promotion/filtering rates} and \emph{agreement with human judgments} across model scales affect ARHN’s behavior and performance.
\item On the BEIR benchmark, we show that dense retrievers fine-tuned on ARHN-refined data achieve the best performance among data-relabeling methods.

\end{itemize}

\begin{figure*}[t]
    \centering
    \includegraphics[width=0.95\textwidth]{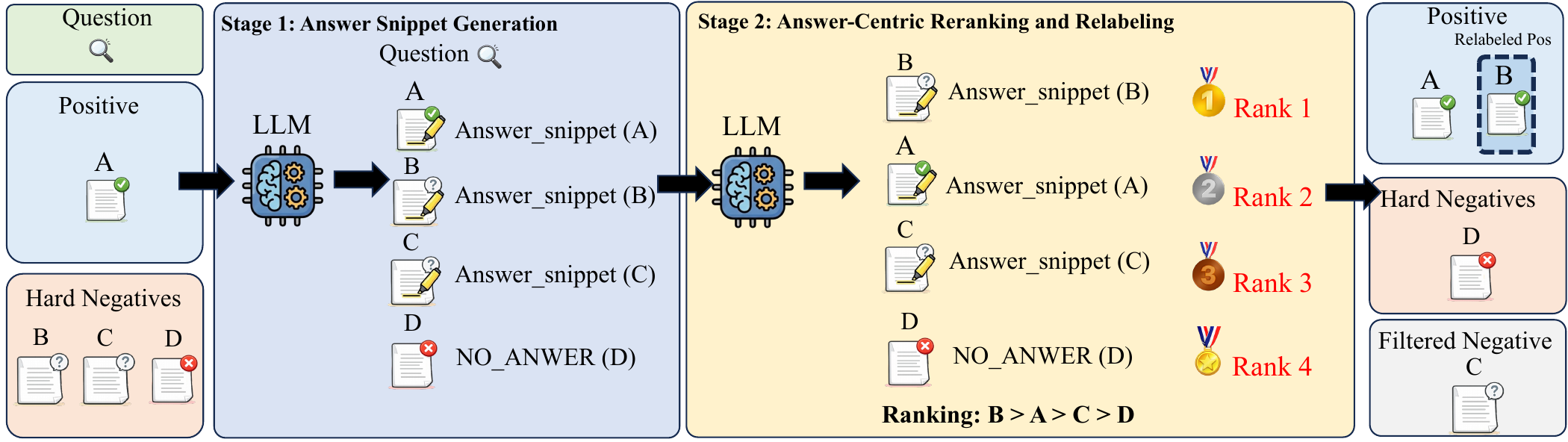}
    \caption{\textbf{Overview of the two-stage ARHN pipeline.}
Given a query $q$, the original training instance consists of a positive document $A$ and hard negative documents $\{B,C,D\}$.
In \textbf{Stage~1}, an open-source LLM extracts \emph{an answer snippet} for the query from the \emph{document text}; if no supporting evidence is found, it outputs \texttt{NO\_ANSWER} (e.g., $D$).
In \textbf{Stage~2}, the extracted snippets are compared in a listwise manner to produce an ordering (e.g., $B>A>C>D$), which is then used to reconstruct labels.}
    \label{fig:arhn_pipeline}
\end{figure*}

\section{Related Work}

\subsection{False Negatives in Retrieval Task}

False negatives—passages that are relevant but treated as negatives due to incomplete annotations and pooling bias—are a persistent source of noise in dense retrieval. When repeatedly sampled during contrastive training, such mislabeled negatives can yield conflicting gradients, pushing representations away from genuinely relevant content and degrading both effectiveness and training stability.

Prior work has proposed several complementary remedies. A teacher-guided approach uses cross-encoder relevance signals to refine bi-encoder supervision, alleviating the adverse impact of false negatives arising from hard negative mining and weak labels \cite{qu2021rocketqa}. Another line improves robustness at the objective level by introducing confidence-aware regularization, reducing sensitivity to potentially corrupted negative sets without requiring explicit relabeling or perfectly curated data \cite{wang2024mitigating}.

Finally, mining-oriented methods address false negatives by improving negative selection itself. Positive-aware mining and systematic filtering of overly hard negatives reduce the chance of sampling passages that are semantically close to positives, leading to more reliable hard negative sets and improved retriever training \cite{moreira2024nv}.

Overall, existing studies treat false negatives as a key bottleneck in retrieval-style learning and mitigate their impact through teacher-based supervision \cite{qu2021rocketqa}, confidence-robust objectives \cite{wang2024mitigating}, uncertainty-aware supervision modeling \cite{ni2021mitigating}, and improved mining and filtering strategies \cite{moreira2024nv}.
\label{sec:related_work}

\subsection{LLM-Assisted Relabeling and Data Curation for Dense Retrieval}

Recent work has increasingly leveraged large language models (LLMs) and strong teacher models to improve dense retrieval training data, motivated by the observation that large-scale collections often contain noisy supervision—most notably false negatives produced by imperfect relevance labels and aggressive hard-negative mining. Instead of treating labels as fixed, these methods use LLMs either to (i) generate new supervision (e.g., synthetic queries or pairs) and build data pipelines, or to (ii) relabel/curate positives and hard negatives by reassessing relevance with stronger semantic judgments.

A representative distillation-style pipeline is Gecko, which constructs synthetic training pairs and then retrieves candidate passages to form positive and hard-negative sets that are subsequently re-labeled using an LLM, producing higher-quality supervision for training text embeddings \cite{lee2024gecko}. While this approach centers on synthetic query generation, its key contribution is an LLM-in-the-loop mechanism for refining positive and hard-negative assignments using teacher-level relevance assessment.

Another line of work focuses on LLM-driven data synthesis for task adaptation. Promptagator demonstrates that a dense retriever can be adapted to new tasks with only a handful of examples by prompting an LLM to generate task-specific queries and training pairs, effectively turning few-shot supervision into scalable retrieval training data \cite{dai2022promptagator}. Similarly, InPars uses LLMs as query generators to augment information retrieval datasets, producing synthetic queries paired with passages to expand training signals beyond limited labeled data \cite{bonifacio2022inpars}. These methods primarily address data scarcity and domain transfer, but they also indirectly mitigate label incompleteness by increasing coverage of plausible query–passage relationships.

Building on synthetic supervision, Noisy self-training with synthetic queries explicitly treats LLM-generated data as noisy and designs a self-training/relabeling loop to iteratively refine the retriever using its own retrieval outputs under a noise-aware regime \cite{jiang2023noisy}. By acknowledging that synthetic labels are imperfect, this direction aligns with broader efforts to make retrieval training robust to mislabeled negatives and uncertain supervision.

Among prior work on refining large-scale training collections, RLHN (ReLabeling Hard Negatives) shows that false negatives and label noise within mined hard-negative sets are a major source of performance degradation for dense retrievers \cite{thakur2025hard}. RLHN points out that false negatives and label noise in hard-negative sets can degrade dense retriever training and proposes an LLM-based relabeling framework that re-evaluates mined negatives to correct mislabeled instances. In RLHN, GPT-4o-mini and GPT-4o are used as the LLMs for the relabeling pipeline.

Overall, these studies illustrate a shift toward LLM-assisted supervision in dense retrieval: from generating synthetic data pipelines for adaptation \cite{dai2022promptagator,bonifacio2022inpars}, to iterative refinement under noise \cite{jiang2023noisy}, and to explicit relabeling/curation of hard negatives and false negatives using LLM-based relevance judgments \cite{lee2024gecko,thakur2025hard}.


\begin{figure}[t]
  \centering
  \includegraphics[width=0.99\columnwidth]{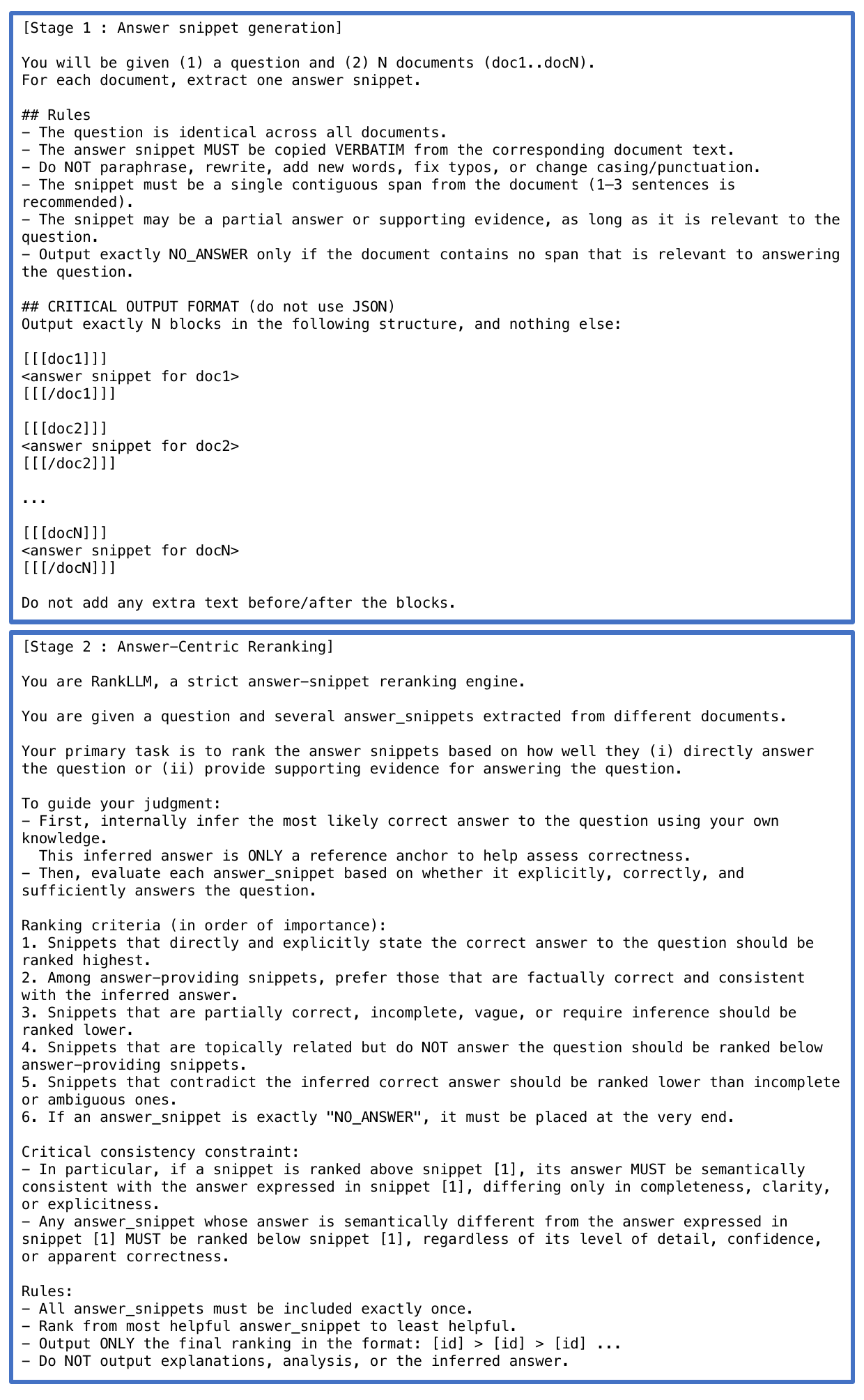}
  \caption{Stage~1--2 prompts of ARHN. In Stage~1, the LLM extracts \emph{an answer snippet} from each query--document pair as a verbatim contiguous span (or outputs \texttt{NO\_ANSWER} when no relevant evidence exists) to verify whether a candidate contains answer-supporting evidence. In Stage~2, the LLM performs answer-centric listwise reranking over the extracted snippets by producing a total order based on how well each snippet answers or supports the query; \texttt{snippet[1]} corresponds to the positive document's \emph{an answer snippet}.}

  \label{fig:stage1_stage2_prompts}
\end{figure}

\section{Method}
\label{sec:method}
In this section, we describe the detailed architecture of the proposed ARHN (Answer-Centric Relabeling of Hard Negatives) framework. Figure 2 provides an overview of the entire ARHN pipeline. 

\subsection{Stage 1: Answer Snippet Generation}
\label{sec:stage1}

The first stage of ARHN determines whether each query--document pair $(q, d)$ actually contains \emph{answer information for the query}. To this end, we use a large language model (LLM) to extract an \emph{answer snippet} from the document that corresponds to the query.
\subsubsection{LLM Input Construction}

For each training instance, the input to the LLM consists of a single query $q$ and a total of $N+1$ documents associated with the query. Specifically, the input includes the following three components.

\begin{itemize}
\item Query $q$
\item Positive document set: $\mathcal{D}^{+}=\{d^{+}\}$
\item Hard negative document set: $\mathcal{D}^{-}=\{d_1^{-}, d_2^{-}, \dots, d_N^{-}\}$
\end{itemize}

That is, for a given query $q$, we simultaneously consider a total of $N+1$ candidate documents, consisting of the positive document $d^{+}$ and $N$ hard negative documents, and prompt the LLM to \emph{independently} judge whether each document contains an answer. Formally, the set of input documents to the LLM is defined as:
\begin{equation}
\mathcal{D}(q) = \mathcal{D}^{+} \cup \mathcal{D}^{-}, \quad |\mathcal{D}(q)| = N + 1.
\end{equation}

Let $\mathcal{D}(q)$ denote the candidate document set for each query $q$. For every $d \in \mathcal{D}(q)$, the LLM extracts \emph{an answer snippet} from the document as follows:
\begin{equation}
\forall d \in \mathcal{D}(q), \quad a = f_{\text{LLM}}(q, d).
\end{equation}

Here, $a$ denotes \emph{an answer snippet}, which must be a sentence or phrase \emph{explicitly contained} in document $d$ in response to query $q$. Importantly, rather than generating a new answer, the LLM is constrained to extract. \emph{an answer snippet} from the document that corresponds to the query. This design restricts answer generation, thereby minimizing hallucinations and encouraging the model to focus on determining \emph{whether an answer to query $q$ exists within} document $d$. Conversely, if the document contains no answer to the query, the LLM outputs a special token indicating the absence of an answer:
\begin{equation}
a = \texttt{NO\_ANSWER}.
\end{equation}
Therefore, the output of Stage~1 for each document is restricted to either \emph{an answer snippet} or \texttt{NO\_ANSWER}. This output serves as a key input signal in Stage~2 for listwise reranking by comparing answer snippet candidates and for identifying false negatives. In particular, when a hard negative document yields $a \neq \texttt{NO\_ANSWER}$, the document is regarded as a potential false-negative candidate because it may contain answer evidence.

When the number of hard negatives is $N$, the output of Stage~1 consists of $(N{+}1)$ tuples:
\begin{equation}
\left\{
(q, d^{+}, a^{+}),
(q, d_1^{-}, a_1^{-}),
\dots,
(q, d_N^{-}, a_N^{-})
\right\}.
\end{equation}
Here, $a^{+}$ denotes the \emph{answer snippet} extracted from the positive document, and each $a_i^{-}$ is either the \emph{answer snippet} extracted from a hard-negative document or \texttt{NO\_ANSWER}. This output serves as input to the subsequent answer-centric ranking and label reconstruction steps in Stage~2.

\subsection{Stage 2: Answer-Centric Reranking and Relabeling}
\label{sec:stage2}

The second stage of ARHN takes as input the \emph{answer snippets} extracted in Stage~1, reranks candidate documents according to their \emph{relative answer correctness} with respect to the query $q$, and incorporates the reranking outcomes into label reconstruction through an \emph{answer-centric reranking and relabeling} procedure. The key idea is not to detect false negatives hidden among hard negatives via absolute score-based grading, but instead to directly obtain from an LLM and leverage a \emph{total order (ordering) among answer snippets}.
\subsubsection{Listwise Reranking via LLM Prompting}

Given a query $q$ and a snippet set $\mathcal{A}(q)$, the LLM outputs a ranking string that sorts answer snippets in \emph{descending} order of answer correctness:
\begin{equation}
[r_1] > [r_2] > \dots > [r_{N+1}],
\end{equation}
Here, $r_1$ denotes the identifier (id) of the snippet that provides the most direct and explicit answer to $q$. We can define this procedure functionally as follows.
\begin{equation}
f_{\text{LLM}}(q, \mathcal{A}(q)) = [r_1, r_2, \dots, r_{N+1}],
\end{equation}
$r_t$ indicates the snippet id at rank position $t$. We further define the rank of a particular snippet id $i$ as follows.
\begin{equation}
\mathrm{rank}_q(i) \triangleq t \quad \text{s.t.} \quad r_t = i,
\end{equation}
That is, a smaller $\mathrm{rank}_q(i)$ indicates that the snippet provides a more explicit answer to the query.
\subsubsection{Rank-Based Answer-Centric Relabeling}

ARHN uses the original positive document $d^{+}$ as an \emph{anchor} to recalibrate the labels of hard-negative documents. Let $i^{+}$ denote the id of the positive snippet. For each hard-negative document $d_i^{-}$, we apply the following rules.

\begin{itemize}
    \item Positive Relabeling
\end{itemize}

\begin{equation}
\mathrm{rank}_q(i) < \mathrm{rank}_q(i^{+})
\;\Rightarrow\;
d_i^{-}\ \text{is promoted to}\ \mathcal{D}^{+}.
\end{equation}

If the snippet from a hard negative is ranked above the positive snippet, we regard the document as a false negative that provides more direct or comparable answer evidence for the query, and thus promote it to the positive set.

\begin{equation}
\begin{aligned}
\mathrm{rank}_q(i) > \mathrm{rank}_q(i^{+}) \;\land\; a_i^{-} \neq \texttt{NO\_ANSWER}
\\
\Rightarrow\; d_i^{-}\ \text{is removed from}\ \mathcal{D}^{-}.
\end{aligned}
\end{equation}

Candidates ranked below the positive yet still containing \emph{an answer snippet} are treated as borderline samples: they may be partially relevant, but their answer correctness is difficult to ascertain. Forcing them to be negatives risks suppressing useful signals, while treating them as positives may increase label noise; therefore, we exclude them from the training dataset.

\begin{itemize}
  \item True Hard Negative Retention
\end{itemize}
\begin{equation}
a_i^{-}=\texttt{NO\_ANSWER}
\;\Rightarrow\;
d_i^{-}\ \text{is kept in}\ \mathcal{D}^{-}.
\end{equation}

Hard negatives for which Stage~1 fails to extract \emph{an answer snippet} provide no content that directly answers the query or supports an answer; therefore, we retain them as true hard negatives (true negatives) and use them for contrastive learning.

\newcommand{\oodmark}{\textsuperscript{\large *}}
\begin{table*}[t]
\centering
\small
\setlength{\tabcolsep}{4pt}
\renewcommand{\arraystretch}{1.05}
\caption{nDCG@10 on 16 BEIR datasets for E5-base and LG-ANNA-Embedding (Mistral-7B), comparing No Refinement, baselines from prior work, and ARHN variants (Filter/Relabel/R+F). Datasets marked with \oodmark correspond to the 7 out-of-domain (OOD) datasets not seen during training; we report the average over all 16 datasets (Avg. 16) and over the 7 OOD datasets (Avg. 7) at the bottom. Best results for each encoder are in bold.}

\resizebox{\textwidth}{!}{%
\begin{tabular}{l|cccccc|cccc}
\toprule
\textbf{BEIR Dataset}
& \multicolumn{6}{c|}{\textbf{E5 (base)}}
& \multicolumn{4}{c}{\textbf{LG-ANNA-Embedding (Mistral-7B)}} \\
\cmidrule(lr){2-7}\cmidrule(lr){8-11}
& \textbf{No Refinement} & \textbf{TopK-PercPos} & \textbf{RLHN} & \textbf{ARHN(Filter)} & \textbf{ARHN(Relabel)} & \textbf{ARHN(R+F)}
& \textbf{No Refinement} & \textbf{ARHN(Filter)} & \textbf{ARHN(Relabel)} & \textbf{ARHN(R+F)} \\
\midrule
BioASQ\textsuperscript{*}         & 0.378 & 0.375 & 0.394 & 0.381 & 0.385 & 0.401 & 0.413 & 0.420 & 0.422 & 0.409 \\
Robust04\textsuperscript{*}       & 0.442 & 0.451 & 0.497 & 0.453 & 0.452 & 0.479 & 0.475 & 0.474 & 0.482 & 0.493 \\
Signal-1M (RT)\textsuperscript{*} & 0.275 & 0.272 & 0.274 & 0.279 & 0.275 & 0.281 & 0.296 & 0.301 & 0.313 & 0.324 \\
TREC-NEWS\textsuperscript{*}      & 0.465 & 0.466 & 0.484 & 0.469 & 0.466 & 0.473 & 0.489 & 0.483 & 0.483 & 0.491 \\
Touch\'e-2020\textsuperscript{*}  & 0.256 & 0.286 & 0.266 & 0.251 & 0.271 & 0.308 & 0.304 & 0.291 & 0.295 & 0.302 \\
TREC-COVID\textsuperscript{*}     & 0.783 & 0.789 & 0.809 & 0.785 & 0.793 & 0.798 & 0.881 & 0.885 & 0.891 & 0.894 \\
NFCorpus\textsuperscript{*}       & 0.378 & 0.377 & 0.390 & 0.374 & 0.380 & 0.381 & 0.393 & 0.402 & 0.401 & 0.414 \\
NQ                                 & 0.595 & 0.601 & 0.591 & 0.592 & 0.592 & 0.612 & 0.641 & 0.643 & 0.638 & 0.641 \\
HotpotQA                           & 0.737 & 0.734 & 0.735 & 0.740 & 0.736 & 0.739 & 0.761 & 0.764 & 0.759 & 0.763 \\
FiQA-2018                          & 0.439 & 0.434 & 0.448 & 0.441 & 0.440 & 0.434 & 0.571 & 0.573 & 0.586 & 0.595 \\
ArguAna                            & 0.701 & 0.697 & 0.692 & 0.700 & 0.706 & 0.719 & 0.724 & 0.729 & 0.721 & 0.719 \\
DBPedia                            & 0.438 & 0.444 & 0.447 & 0.439 & 0.437 & 0.442 & 0.494 & 0.511 & 0.504 & 0.514 \\
SCIDOCS                            & 0.242 & 0.243 & 0.242 & 0.243 & 0.243 & 0.252 & 0.252 & 0.241 & 0.243 & 0.240 \\
FEVER                              & 0.878 & 0.878 & 0.871 & 0.880 & 0.876 & 0.879 & 0.881 & 0.879 & 0.881 & 0.883 \\
Climate-FEVER                      & 0.391 & 0.386 & 0.367 & 0.393 & 0.385 & 0.391 & 0.395 & 0.397 & 0.421 & 0.419 \\
SciFact                            & 0.735 & 0.735 & 0.740 & 0.739 & 0.731 & 0.741 & 0.769 & 0.763 & 0.771 & 0.769 \\
\midrule
Avg. 16 (All)                      & 0.508 & 0.511 & 0.515 & 0.510 & 0.511 & \textbf{0.521} & 0.546 & 0.547 & 0.551 & \textbf{0.554} \\
Avg. 7 (OOD)                       & 0.425 & 0.431 & 0.445 & 0.427 & 0.432 & \textbf{0.446} & 0.464 & 0.465 & 0.470 & \textbf{0.475} \\
\bottomrule
\end{tabular}%
}
\label{tab:beir_e5_mistral}
\end{table*}

\section{Experimental Setting}
\label{sec:exp}

\subsection{Training Data and Refinement Setup}
We apply ARHN to refine the original BGE training collection \cite{li2024making}. The BGE collection includes multi-task training data gathered from diverse tasks, including retrieval, clustering , and classification. Although BGE contains a large number of query--passage pairs (approximately 1.6M) collected from various sources, the RLHN study \cite{thakur2025hard} reports that some datasets can negatively affect model effectiveness and that pruning parts of the BGE collection can improve overall performance. In particular, \cite{thakur2025hard} shows that training on a specific subset of seven datasets (MS MARCO, HotpotQA, NQ, FEVER, SciDocsRR, FiQA-2018, and ArguAna) yields better overall performance than using the full corpus. Following \cite{thakur2025hard}, we keep the same seven-dataset configuration and apply ARHN to improve the quality of hard negatives.

For the LLMs used in Stage~1 and Stage~2, we conduct experiments with Qwen3-8B, Qwen3-14B, and Qwen3-32B, and report the final results using Qwen3-32B. Each prompt includes one positive document and up to $N=10$ hard negatives per query. We ran LLM inference for both stages on 16 NVIDIA A100 GPUs.

\subsection{Base Retriever Models}
We evaluate ARHN using two dense retrieval models.  As a BERT-based bi-encoder backbone \cite{devlin2019bert}, we use E5-base \cite{wang2024improving}, which has approximately 110M parameters, uses a 12-layer Transformer with 768-dimensional embeddings, and produces sentence representations via mean pooling.

To further assess whether ARHN generalizes to LLM-based embedding models, we also include LG-ANNA-Embedding \cite{choi2025lg}, a Mistral-7B-based general-purpose text embedder \cite{chaplot2023albert}. LG-ANNA-Embedding adopts an instruction-following framework that combines context-aware prompting, soft labeling, and adaptive-margin hard-negative mining, and it achieves strong performance on MTEB (English, v2) according to the Borda score \cite{choi2025lg}.

\subsection{Training Details}
All models used in our experiments are optimized with the InfoNCE loss \cite{izacard2021unsupervised}. For each query, a training instance consists of one positive passage and seven hard negatives, and we also leverage in-batch negatives. The global batch size is set to 128.

We run each experiment with multiple random seeds and report the mean performance across runs. We fine-tune E5-base for five epochs with a learning rate of $2 \times 10^{-5}$. For LG-ANNA-Embedding, we adopt parameter-efficient fine-tuning with Low-Rank Adaptation (LoRA) \cite{hu2022lora} and fine-tune the model for two epochs with a learning rate of $1 \times 10^{-4}$. Both E5-base and the Mistral-7B-based model were fine-tuned on NVIDIA A100 GPUs.

\subsection{Evaluation Datasets and Metrics}
We evaluate models fine-tuned on ARHN-reconstructed data on the BEIR benchmark \cite{thakur2021beir}. We use nDCG@10 as the standard retrieval metric. To ensure comparability with prior work, we report results on 16 of the 18 BEIR datasets, excluding Quora and CQADupStack.

\subsection{Comparison Methods and ARHN Variants}
To quantitatively assess the effect of ARHN on hard-negative refinement, we compare (i) a baseline trained on the default training set, (ii) a data-refinement strategy from prior work, and (iii) ARHN variants that ablate individual components. Here, the default training set is constructed by retaining retrieval data from the BGE training collection and then pruning it to the seven datasets that contribute most to performance, yielding approximately 680K query--passage pairs.

\paragraph{(0) No Refinement.}
The model is fine-tuned on the default training set without any additional refinement. This serves as the primary reference point, measuring the performance attainable without data reconstruction and enabling us to quantify \emph{data-level} improvements from subsequent methods.

\paragraph{(1) TopK-PercPos.}
This baseline follows the hard-negative mining procedure proposed in NV-Retriever \cite{moreira2024nv}. Specifically, for each query, we score negative candidates using the bge-reranker-v2-gemma reranker and then apply TopK-PercPos (top-95\%) sampling to construct the hard-negative set used for training.

\paragraph{(2) RLHN (ReLabeling Hard Negatives).}
RLHN points out that false negatives and label noise in hard-negative sets can degrade dense retriever training and proposes an LLM-based relabeling framework that re-evaluates mined negatives to correct mislabeled instances. In RLHN, GPT-4o-mini and GPT-4o are used as the LLMs for the relabeling pipeline.

\paragraph{(3) ARHN (Filter / Relabel / R+F)}
ARHN extracts \emph{answer evidence (answer snippets)} from each document and then reconstructs the training data by re-evaluating hard negatives in a listwise manner based on the extracted evidence. This process mitigates the impact of false negatives in hard-negative sets. We compare two component variants and their combined setting: (i) \textbf{ARHN(Filter)} filters out negatives whose answer snippets are ranked lower than the positive document's answer snippet in the listwise ranking, even when \emph{an answer snippet} is produced for the negative. (ii) \textbf{ARHN(Relabel)} promotes a negative to a positive (relabeling) when its answer snippet is ranked higher than the positive document's answer snippet in the listwise ranking. (iii) \textbf{ARHN(R+F)} applies both relabeling and filtering and serves as the final variant.

\section{Experimental Results}


\begin{table}[t]
\centering
\caption{nDCG@10 comparison between PRHN and ARHN  on 16 BEIR datasets, using an E5-base retriever fine-tuned on the corresponding refined training data. PRHN performs the Stage~2 listwise ranking directly on the original passages, whereas ARHN first generates \emph{an answer snippet} for each passage in Stage~1 and then conducts Stage~2 ranking conditioned on the extracted snippets.}

\label{tab:prhn_arhn}
\setlength{\tabcolsep}{4pt} 
\renewcommand{\arraystretch}{1.05}

\begin{tabularx}{\columnwidth}{l
  S[table-format=1.3]
  S[table-format=1.3]}
\toprule
\textbf{Dataset} &
\multicolumn{1}{c}{\makecell{\textbf{PRHN}\\\scriptsize \textbf{P}assage-Centric \textbf{R}elabeling\\\scriptsize of \textbf{H}ard \textbf{N}egatives}} &
\multicolumn{1}{c}{\makecell{\textbf{ARHN}\\\scriptsize \textbf{A}nswer-Centric \textbf{R}elabeling\\\scriptsize of \textbf{H}ard \textbf{N}egatives}} \\
\midrule
BioASQ*         & 0.391 & 0.401 \\
Robust04*       & 0.452 & 0.479 \\
Signal-1M (RT)* & 0.285 & 0.281 \\
TREC-NEWS*      & 0.464 & 0.473 \\
Touche2020*     & 0.311 & 0.308 \\
TREC-COVID*     & 0.801 & 0.798 \\
NFCorpus*       & 0.379 & 0.381 \\
NQ              & 0.611 & 0.612 \\
HotpotQA        & 0.736 & 0.739 \\
FiQA-2018       & 0.432 & 0.434 \\
ArguAna         & 0.721 & 0.719 \\
DBPedia         & 0.439 & 0.442 \\
SCIDOCS         & 0.242 & 0.252 \\
FEVER           & 0.863 & 0.879 \\
Climate-FEVER   & 0.381 & 0.391 \\
SciFact         & 0.747 & 0.741 \\
\midrule
Avg.\ 16 (All)  & 0.516 & 0.521 \\
Avg.\ 7 (OOD)   & 0.440 & 0.446 \\
\bottomrule
\end{tabularx}

\end{table}

\begin{figure}[t]
  \centering
  \includegraphics[width=0.85\columnwidth]{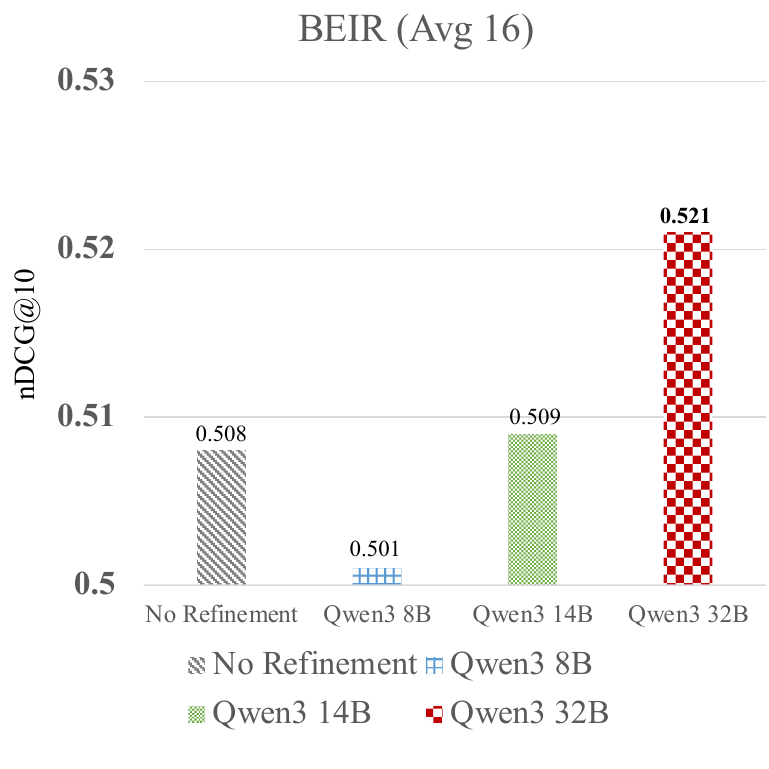}
  \caption{Effect of LLM scale used in ARHN labeling (Stage~1--2) on retrieval performance. The plot reports nDCG@10 on BEIR (Avg.\ 16) for an E5-base retriever fine-tuned on ARHN(R+F)-refined data, comparing No Refinement with Qwen3-8B/14B/32B. Performance improves with larger LLMs, and Qwen3-32B achieves the best nDCG@10.
}
  \label{fig:qwen3_scale_beir}
\end{figure}

\subsection{Results on the BEIR Benchmark}
\label{sec:exp_results_beir_kr}

Table~\ref{tab:beir_e5_mistral} reports nDCG@10 on 16 BEIR datasets and compares (i) training without refinement (No Refinement), (ii) baselines from prior work (TopK-PercPos, RLHN), and (iii) three variants of the proposed ARHN (Filter/Relabel/Relabel+Filter). The table also presents results for two retriever models, E5-base and LG-ANNA-Embedding (Mistral-7B), to examine whether the same refinement strategy generalizes across different retrieval models.

\subsubsection{Overall improvements: consistent gains from ARHN(R+F)}
With the E5-base backbone, ARHN(R+F) achieves an Avg.\ 16 (All) score of 0.521, improving over No Refinement (0.508) by 1.3 nDCG@10 points. ARHN(R+F) also outperforms TopK-PercPos (0.511) and RLHN (0.515), indicating that the proposed relabeling and filtering operations more effectively mitigate false negatives and label noise in the hard-negative set.

With LG-ANNA-Embedding (Mistral-7B), ARHN(R+F) attains the best Avg.\ 16 (All) score of 0.554, improving over No Refinement (0.546) by 0.8 points. These results indicate that ARHN does not depend on a specific backbone and tends to improve different retrieval models by enhancing training data quality.

\subsubsection{OOD generalization: benefits of mitigating label noise}
On the seven OOD datasets marked with \textsuperscript{*} (i.e., domains unseen during training), E5-base improves from 0.425 to 0.446 on Avg.\ 7 (OOD), a gain of 2.1 points. The larger improvement on Avg.\ 7 (OOD) than on Avg.\ 16 suggests that relabeling and filtering false negatives are especially beneficial for OOD generalization.
LG-ANNA-Embedding (Mistral-7B) shows a similar pattern: Avg.\ 7 (OOD) increases from 0.464 to 0.475 (1.1 points), supporting the conclusion that ARHN strengthens OOD generalization across different retrieval models.

\subsubsection{Relabeling vs.\ filtering: complementarity and synergy}
A comparison of ARHN variants shows that combining filtering and relabeling yields larger gains than applying either operation alone. With E5-base, ARHN(Filter) reaches 0.510 and ARHN(Relabel) reaches 0.511, which are only modest improvements over No Refinement (0.508), whereas ARHN(R+F) increases performance to 0.521. This pattern suggests that relabeling and filtering are complementary: relabeling corrects mislabeled hard negatives, while filtering removes ambiguous negatives. Applying both yields a synergistic effect.

(1) Relabeling promotes false negatives---hard negatives that actually contain \emph{an answer snippet}---to positives. This correction strengthens the supervision signal and increases the \emph{diversity} of positive examples used for training.

Relabeling can also reduce the impact of \emph{false positives} in the original training data (i.e., documents labeled as positives but lacking a sufficient answer snippet), because newly identified high-quality positives can compensate for weak or noisy supervision.

In addition, in our two-stage LLM labeling pipeline, the model extracts {an answer snippet} after truncating each input document to the retriever's \texttt{max\_seq\_len}. If the answer snippet in an original positive lies beyond \texttt{max\_seq\_len}, the retriever cannot observe it during training, whereas some hard negatives may contain an explicit answer snippet within \texttt{max\_seq\_len}. Promoting such negatives to positives makes an observable answer snippet available within \texttt{max\_seq\_len} and thus provides a stronger and more direct learning signal.

(2) Filtering removes borderline negatives, such as partially relevant documents whose extracted answer snippet provides only incomplete evidence, thereby reducing the risk of training the model to aggressively push away partially relevant documents.

\subsubsection{Comparison with RLHN: API-based refinement vs.\ open-source-LLM refinement}RLHN uses proprietary LLMs accessed via an API (GPT-4o-mini and GPT-4o) to refine hard negatives, whereas ARHN performs labeling---answer snippet extraction and listwise reranking---using an open-source LLM such as Qwen3-32B.

Table~\ref{tab:beir_e5_mistral} shows that ARHN(R+F) achieves 0.521 with E5-base, outperforming RLHN (0.515), and yields a higher OOD average. These results suggest that a sufficiently capable open-source LLM can improve training data quality and deliver gains comparable to, or larger than, those obtained with API-based refinement. Open-source refinement also offers practical benefits in terms of cost and reproducibility.
\begin{table}[t]
\centering
\caption{Cohen's $\kappa$ between LLM labels and human judgments on 500 query--negative pairs.}
\label{tab:kappa_qwen}
\small
\setlength{\tabcolsep}{6pt}
\renewcommand{\arraystretch}{1.1}

\begin{tabular}{lccc}
\toprule
\textbf{Metric} 
& {\footnotesize\textbf{Qwen3-8B}}
& {\footnotesize\textbf{Qwen3-14B}}
& {\footnotesize\textbf{Qwen3-32B}} \\
\midrule
Cohen's Kappa ($\kappa$) & 0.312 & 0.341 & 0.373 \\
\bottomrule
\end{tabular}
\end{table}

\begin{table}[t]
\centering
\small
\setlength{\tabcolsep}{6pt}
\renewcommand{\arraystretch}{1.15}
\caption{Refinement statistics of ARHN(R+F) across LLM scales. For each LLM (Qwen3-8B/14B/32B) with $N{=}10$ hard negatives per query, we report the average number of negatives relabeled as positives (Relabeled Pos.) and the average number of negatives removed (Filtered Neg.) per query.}

\begin{tabular}{lccc}
\toprule
\textbf{LLM} & \textbf{$N$} & \textbf{Relabeled Pos.} & \textbf{Filtered Neg.} \\
\midrule
Qwen3 8B  & 10 & 3.1   & 3.9    \\
Qwen3 14B & 10 & 2.3   & 3.4   \\
Qwen3 32B & 10 & 1.6  & 2.2   \\
\bottomrule
\end{tabular}
\label{tab:hn_restruct_stats}
\end{table}

\begin{table*}[t]
\centering
\small
\setlength{\tabcolsep}{8pt}
\renewcommand{\arraystretch}{1.25}
\caption{Examples of label noise in retrieval training (FiQA-2018, NQ, MS MARCO). Highlighted text denotes \emph{an answer snippet}. We contrast labeled positives with answer-bearing negatives, illustrating complementary evidence, more specific answers, true-answer false negatives, and contaminated negatives with overlapping answer snippets.
}

\begin{tabularx}{\textwidth}{
    >{\raggedright\arraybackslash}p{0.06\textwidth}
    >{\raggedright\arraybackslash}p{0.18\textwidth}
    >{\raggedright\arraybackslash}X
    >{\raggedright\arraybackslash}X
}
\toprule
\textbf{Data} & \textbf{Query} & \textbf{Positive Passages} & \textbf{Relabeled Positive (False Negatives)} \\
\midrule

\textbf{FIQA-2018}
&
How to Deduct Family Health Care Premiums Under Side Business
&
\textbf{Positive1 :} [\dots] You received wages in 2011 from an S corporation in which you were a more-than-2\% shareholder. \hl{Health insurance premiums paid or reimbursed by the S corporation are shown as wages on Form W-2. The insurance plan must be established under your business. Your personal services must have been a material income-producing factor in the business. If you are filing Schedule C, C-EZ, or F, the policy can be either in your name or in the name of the business.}
&
\textbf{Negative1 :} [\dots] So the self-employed person has to pay both the employer's share as well as the employee's share of Social Security and Medicare taxes on that money. \hl{Health insurance premiums can be deducted on Line 29 of Form 1040 but only for those months during which the Schedule C filer is neither covered nor eligible to be covered by a subsidized health insurance plan maintained by an employer of the self-employed person (whose self-employment might be a sideline) or the self-employed person's spouse.}  [\dots] 
\\
\midrule

\textbf{NQ}
&
when was the united states pledge of allegiance adopted
&
\textbf{Positive1 :} [\dots] The form of the pledge used today was largely devised by Francis Bellamy in 1892, and formally adopted by Congress as the pledge in \hl{1942}. The official name of "The Pledge of Allegiance" was adopted in 1945.
&
\textbf{Negative1 :} [\dots] Congress officially recognized the Pledge for the first time, in the following form, on \hl{June 22, 1942}: Louis Albert Bowman, an attorney from Illinois, was the first to suggest the addition of "under God" to the pledge [\dots] 
\\
\midrule
\textbf{Data} & \textbf{Query} & \textbf{Positive Passages} & \textbf{Filtered Negative} \\
\midrule
\textbf{MS marco} & meds that can cause irregular heartbeat &\textbf{Positive1 :}  [\dots] Always advise your doctor of any medications or treatments you are using, including prescription, over-the-counter, supplements, herbal or alternative treatments. \hl{1 Aldazine. 2 Amphetamine Sulfate. 3 Anatensol}. &\textbf{Negative1 :}  \hl{Cardiac Side Effects of Lithium Lithium may cause arrhythmias, or irregular heartbeat, throughout the course of therapy.} If the patient experiences heart palpitations or uneven heart beat, he should seek medical care right away.  
 \\
\midrule
\textbf{MS marco} & what are normal numbers for glaucoma &\textbf{Positive1 :} Glaucoma and Eye Pressure: Q\&A A: \hl{Normal pressure in the eye is between 12 and 21mm Hg}. Some patients are fortunate in that their optic nerves can tolerate pressures outside this range.[\dots] & \textbf{Negative1 :} [\dots] Eye pressure, called intraocular pressure (IOP), is measured in millimeters of mercury (mm Hg).\hl{Normal eye pressure ranges from 10-21 mm Hg}[\dots] 
\newline 
\textbf{Negative2 :}  What Is the Normal Range for Eye Pressure? Glaucoma is an eye condition that is caused by increased intraocular pressure. \hl{Normal range for eye pressure is between 10- to 21-mm HG.} [\dots] \newline 

\\

\bottomrule
\end{tabularx}
\label{tab:qual_example_nq}
\end{table*}

\subsection{PRHN vs. ARHN: Passage-Centric vs. Answer-Centric Relabeling}
\label{sec:prhn_arhn}
Table~\ref{tab:prhn_arhn} compares PRHN, which performs Stage~2 listwise ranking directly on the original passages, with ARHN, which first extracts \emph{an answer snippet} from each passage in Stage~1 and then performs Stage~2 listwise ranking conditioned on the extracted snippets. Both methods apply \emph{relabeling} and \emph{filtering} to hard negatives, but they differ in the input granularity used for decision-making: PRHN evaluates candidates at the passage level, whereas ARHN bases its ranking on extracted \emph{answer snippets}.

ARHN achieves higher overall performance than PRHN. On Avg.\ 16 (All), ARHN improves nDCG@10 from 0.516 to 0.521 (+0.005). ARHN also improves Avg.\ 7 (OOD) from 0.440 to 0.446, suggesting that answer-centric signals can benefit out-of-domain generalization.

This gap is consistent with the difficulty of Stage~2 ranking. In Stage~2, the model ranks candidates based on both how well they support the query and whether they provide a correct answer, yet positives and hard negatives often share similar lexical and contextual cues, making passage-level comparison challenging. ARHN reduces this ambiguity by ranking \emph{an answer snippet} rather than the full passage, which removes shared background context and focuses the comparison on \emph{an answer snippet}. In addition, Stage~1 emits a special token, \texttt{NO\_ANSWER}, for passages that lack \emph{an answer snippet}, which helps Stage~2 more efficiently separate non-evidence candidates.

\subsection{Effect of LLM Scale on ARHN Refinement}
\label{sec:exp_llm_scale}

Figure~\ref{fig:qwen3_scale_beir} examines how the LLM scale used for ARHN(R+F) labeling (Qwen3-8B/14B/32B) affects final retrieval performance (nDCG@10). We observe a positive trend with increasing LLM scale: Qwen3-32B improves Avg.\ 16 (All) from 0.508 to 0.521. This trend is consistent with larger LLMs identifying \emph{an answer snippet} more reliably in Stage~1 and applying relabeling and filtering more consistently in Stage~2, resulting in higher retrieval performance.

\paragraph{Small LLMs can hurt.}
Small LLMs can degrade the quality of refined training data when used for refinement, potentially introducing additional label noise. Qwen3-8B reduced Avg.\ 16 (All) from 0.508 to 0.501. These drops are consistent with refinement errors from weaker LLMs, such as extracting \emph{an answer snippet} from passages without answer evidence, promoting partially relevant passages to positives, or filtering out informative hard negatives. These results show that ARHN does not guarantee improvements; it benefits from refinement only when the labeling model is sufficiently accurate. In practice, ARHN(R+F) yields consistent gains with a strong refinement model such as Qwen3-32B, whereas a smaller model can introduce additional label noise and reduce average performance.

\subsection{Agreement with Human Judgments}
Table~3 summarizes our human validation setup for assessing the reliability of LLM-based labeling. We briefed two human assessors on the false-negative identification task and asked them to independently annotate 500 query--hard-negative pairs. We randomly sampled hard negatives from the training set and constructed the validation set such that each query's hard-negative set contained at least one hard negative that the LLM relabeled as a false negative; we then asked the assessors to identify which hard negatives were false negatives. The assessors did not observe the LLM predictions during annotation. When the two assessors disagreed, they discussed the case and produced a single adjudicated label.

Table~\ref{tab:kappa_qwen} reports Cohen's $\kappa$ between each LLM's predicted labels (Qwen3-8B/14B/32B) and the adjudicated human labels. Agreement increases with model scale: Qwen3-8B achieves $\kappa{=}0.312$, Qwen3-14B achieves $\kappa{=}0.341$, and Qwen3-32B achieves $\kappa{=}0.373$. These results indicate non-trivial agreement with human judgments even on query--hard-negative pairs for which false-negative identification is difficult, and they suggest that larger models more consistently capture cues corresponding to \emph{an answer snippet}.

\subsection{How LLM Scale Shapes Relabeling and Filtering Behavior}

Table~\ref{tab:hn_restruct_stats} helps explain why smaller LLMs can degrade the quality of refined training data. As the LLM size decreases, ARHN(R+F) applies relabeling and filtering more aggressively: Qwen3-8B relabels 3.1 negatives as positives and filters 3.9 negatives per query on average, whereas Qwen3-32B relabels 1.6 and filters 2.2. This pattern is consistent with weaker LLMs making more labeling mistakes, such as extracting \emph{an answer snippet} from passages without answer evidence, promoting partially relevant passages to positives, or filtering out informative hard negatives.

\section{Analysis}

Table~\ref{tab:qual_example_nq} illustrates several labeling failure modes that can materially affect retrieval training. In particular, it shows that (i) multiple passages can be legitimately relevant to the same query, and (ii) some passages labeled as negatives actually contain correct answers (false negatives) or even near-duplicate answer spans (negative contamination). 
\begin{enumerate}
    \item \textbf{Benefit of diverse positives (complementary evidence).}
In the FiQA-2018 example (``How to Deduct Family Health Care Premiums Under Side Business''), the original positive focuses on one subset of requirements (e.g., how the plan is established under the business), whereas the relabeled-positive passage (initially treated as a negative) provides a different but crucial constraint (e.g., eligibility conditions for the Line~29 deduction). These passages are complementary rather than redundant. Allowing multiple positives helps prevent \emph{under-specification}: the model learns a broader notion of what constitutes answer-bearing evidence and can retrieve support that covers multiple subconditions.

    \item \textbf{Benefit of higher specificity and clarity (1942'' $\rightarrow$ June 22, 1942'').}
In the NQ example (when was the United States Pledge of Allegiance adopted''), one passage provides a coarse, year-level answer, whereas another provides a more precise date (June 22, 1942''). If the more specific passage is mislabeled as a negative, training explicitly penalizes retrieval of \emph{better} evidence. Relabeling these passages as positives teaches the model to favor passages that give an exact date rather than a vague year, which matters most for date/time questions.

    \item \textbf{False negatives that are valid answers in practice (the Lithium case).}
In the MS MARCO example (``meds that can cause irregular heartbeat''), a filtered negative states that \emph{Lithium may cause arrhythmias/irregular heartbeat}. Even if the dataset's chosen positive mentions different medications, Lithium remains a valid real-world answer. Treating Lithium-containing passages as negatives teaches the model to suppress legitimate evidence, which can reduce recall of clinically relevant options when the model is deployed in a real-world service and lead to incomplete or misleading outputs.

    \item \textbf{Severe negative contamination: negatives contain (near-)identical answer spans.}
In the MS MARCO example (``what are normal numbers for glaucoma''), the positive passage gives a normal intraocular pressure range, while multiple negatives include essentially the same range (with minor numeric variations such as 10--21 vs.\ 12--21 vs.\ 12--22~mmHg). This creates contradictory supervision, which can ultimately degrade both retrieval quality and training stability.
\end{enumerate}

\section{Conclusion}
Hard-negative mining is essential for training dense retrievers, but mined negatives often contain \emph{false negatives}---answer-bearing passages incorrectly labeled as negatives---which can introduce contradictory supervision and hurt robustness. In this work, we propose \textbf{ARHN},  \emph{an answer-centric} refinement pipeline that uses an open-source LLM to extract \emph{an answer snippet} (or \texttt{NO\_ANSWER}) for each query--document pair and then ranks candidates by direct answerability. 

Experiments on the BEIR benchmark show that jointly applying relabeling and filtering yields the most consistent improvements across retriever models, with particularly strong gains on out-of-domain datasets, indicating that mitigating label noise is crucial for generalization. We also find that open-source LLMs can provide effective and reproducible refinement and that larger models further improve refinement quality. Overall, ARHN offers a practical, scalable approach to cleaning hard-negative supervision and training more robust dense retrieval models by centering decisions on explicit answer evidence.


\bibliographystyle{ACM-Reference-Format}
\bibliography{sample-base}


\end{document}